\documentclass[amsmath,amssymb,preprint,goupaddress,prb,aps,showpacs,showkeys,superscriptaddress]{revtex4}
\usepackage{bm,graphics,subfigure,enumerate,latexsym,comment}
\usepackage[dvips]{epsfig}

\begin{document}
\title{Dynamic rheology of a supercooled polymer melt in non-uniform oscillating flows in rapidly oscillating plates}
\author{Shugo Yasuda
\footnote{Electronic mail: yasuda@sim.u-hyogo.ac.jp}}
\affiliation{
Graduate School of Simulation Studies, University of Hyogo, 
Kobe 650-0047, Japan}
\author{Ryoichi Yamamoto
\footnote{Electronic mail: ryoichi@cheme.kyoto-u.ac.jp}}
\affiliation{
Department of Chemical Engineering, 
Kyoto University, Kyoto 615-8510, Japan, and
CREST, Japan Science and Technology Agency, Kawaguchi 332-0012, Japan
}
\date{\today}
\begin{abstract}
The dynamic rheology of a polymer melt composed of short chains with ten beads between rapidly oscillating plates is investigated for various oscillation frequencies by using the hybrid simulation of the molecular dynamics and computational fluid dynamics.
In the quiescent state, the melt is in a supercooled state, and the stress relaxation function $G(t)$ exhibits a stretched exponential relaxation on the time scale of the $\alpha$ relaxation time $\tau_\alpha$ (the structural relaxation of beads) and then follows the Rouse relaxation function characterized by the Rouse relaxation time $\tau_R$ (the conformational relaxation of polymer chains).
In the rapidly oscillating plates, non-uniform boundary layer flows are generated over the plate due to inertia of the fluid, and the local rheological properties of the melt are spatially varied according to the local flow fields.
The local strain and local strain rate of the melt monotonically decrease with the distance from the plate at each oscillation frequency of the plate, but their dependencies on the oscillation frequency at a fixed distance from the plate vary with the distance.
Far from the plate, the local strain decreases as the oscillation frequency increases such that the dynamic rheology deviates from the linear moduli at the low oscillation frequencies rather than high oscillation frequencies.
On the contrary, near the plate, the local strain rate increases with the oscillation frequency such that the shear thinning is enhanced at high oscillation frequencies.
In close vicinity to the plate, the dynamic viscosity is mostly independent of the oscillation frequency, and the shear thinning behavior becomes similar to that observed in steady shear flows.
We show the diagram of the loss tangent of the melt for different oscillation frequencies and local strain rates.
It is seen that the melt generates three different rheological regimes, i.e., the viscous fluid regime, liquid-like viscoelastic regime, and solid-like viscoelastic regime, according to the oscillation frequency and local strain rate.
Non-linear rheological properties are also investigated by the spectrum analysis and the Lissajous-Bowditch curve.
It is found that the fractional amplitude of the higher harmonics to the linear harmonics is suppressed within the boundary layer due to the non-slip boundary on the oscillating plate.
We also find that the melt exhibits inter-cycle shear thinning between different positions but exhibits intra-cycle shear thickening at a fixed position in the vicinity of the plate.
\end{abstract}


\maketitle

% main text
\section{Introduction}
Glassy polymeric fluids have complicated shear-dependent dynamic rheology.
In steady shear flows, glassy fluids are highly viscous, and shear thinning occurs at sufficiently large shear rates due to chain elongations.\cite{book:92M, book:96S}
In unsteady flows, glassy fluids exhibit elastic behaviors if the characteristic time scale of polymer dynamics is comparable to or larger than that in the flow system.
The viscoelastic property can be measured by the shear moduli, i.e., the storage modulus $G'$ for the elasticity and the loss modulus $G''$ for the viscosity.
The shear moduli can be measured under uniform oscillatory shear flow with a finite shear strain, both experimentally or numerically.
In general non-uniform flows, however, the local rheological properties become heterogeneous depending on the local flow variables.
Thus, the rheological behaviors of glassy polymeric fluids in highly non-uniform flows are so complicated that the theoretical or experimental approaches to this problem are very difficult.
It is also difficult to predict the flow behaviors of such fluids because the reliable constitutive equations are not known in general, although there is an important accumulation of both experimental and theoretical works to construct them.\cite{book:87BAH,book:88L}

In the present paper, we investigate the dynamic rheology of a model polymer melt composed of short chains between rapidly oscillating plates by using the hybrid simulation of molecular dynamics (MD) and computational fluid dynamics (CFD).
The temperature of the melt is so low that the glassy behavior is observed in the stress relaxation function in the quiescent state.
Non-uniform oscillatory shear flows are generated over the oscillating plate at sufficiently large oscillation frequencies due to inertia of the fluid via the term $\rho\partial {\bm v}/\partial t$, which is sometimes called the transient force.
Thus, heterogeneous rheological behaviors arise according to local flow variables.

Oscillatory shear flows under the transient force for the viscoelastic fluid have been investigated by several researchers so far.
Scharg analytically solved the flows of a linear viscoelastic fluid\cite{art:77S}. Dunwoody investigated the inertia effect for a weak non-linear viscoelastic fluid with the K-BKZ model by using a perturbation analysis\cite{art:96D}. Yosick {\it et al}. performed 
numerical analysis for a non-linear viscoelastic fluid with the Berkeley model. 
Ding {\it et al}. investigated the viscous dissipation under the transient force in the temperature field for linear viscoelastic fluids\cite{art:99DGBK}.
In previous works, some constitutive model equations have been used to calculate flow profiles.
In the present hybrid simulation, no constitutive model is required to obtain a local stress.
The local stress is generated by a local MD simulation according to the local flow variables.

In the hybrid simulation method, the macroscopic flows of the melt are calculated by using the CFD scheme; however, instead of using any constitutive equations, the local stresses of the melt are calculated by using molecular dynamic simulations of polymer chains according to the local strain rates.\cite{art:08YY, art:09YY, art:10YY}
The basic idea of the present hybrid simulation method was first proposed by E and Engquist\cite{art:03EE, art:07EELRV}, where the heterogeneous multiscale method (HMM) is presented as a general methodology for the efficient numerical computation of problems with multiscale characteristics.
The HMM has also been applied to the simulation of complex fluids.\cite{art:05RE}
Equation-free multiscale computation was also proposed by Kevrekidis {\it et al}. on the basis of a similar idea and has been applied to various problems.\cite{art:03KGHKRT, art:09KS}
De {\it et al}. have developed a hybrid method, called the scale bridging method, which can correctly reproduce the memory effect of a polymeric liquid and demonstrated non-linear viscoelastic behavior of a polymeric liquid between oscillating plates.\cite{art:06DFSKK}
The multiscale simulation based on a similar idea has been also applied to rarefied gas flows, recently.\cite{art:10KOK}

In what follows, we characterize the linear dynamic rheology of a model polymer melt by using the MD simulation in Sec. II.
Then, we investigate the dynamic rheology of the melt in non-uniform oscillatory shear flows under the transient force between rapidly oscillating plates in Sec. III, where the simulation method, velocity profiles, shear moduli, and LAOS analysis are introduced.
Finally, we summarize the results in Sec. IV.

\section{Linear dynamic rheology of a model polymer melt}
We consider a model polymer melt composed of short chains with ten beads of a uniform density $\rho_0$ and a uniform temperature $T_0$.
The number of bead particles on each chain is represented by $N_{\rm b}$.
Thus, $N_{\rm b}=10$.
All of the bead particles interact with a truncated Lennard-Jones potential defined by\cite{art:90KG},
\begin{equation}\label{eq1}
U_{\rm LJ}(r)=
\left\{
\begin{array}{c c}
4\epsilon\left[
({\sigma}/{r})^{12}
-({\sigma}/{r})^{6}
\right]
+\epsilon & (r\le 2^{1/6}\sigma),\\
0 & ( r> 2^{1/6}\sigma).
\end{array}
\right .
\end{equation}
By using only the repulsive part of the Lennard-Jones potential, we may prevent spatial overlap of the particles.
Consecutive beads on each chain are connected by an anharmonic spring potential,
\begin{equation}\label{eq2}
U_{\rm F}(r)=-\frac{1}{2}k_c R_0^2 \ln
\left[
1-({r}/{R_0})^2
\right],
\end{equation}
where $k_c$=30$\epsilon/\sigma^2$ and $R_0$=$1.5\sigma$.
The temperature of the melt is $k_B T_0/\epsilon$=0.2, where $k_B$ is the Boltzmann constant.
The number density of the bead particles is $\rho_0/m$=1/$\sigma^3$, where $m$ is the mass of the bead particle.
With this number density and temperature, the configuration of the bead particles becomes severely jammed, resulting in a complicated non-Newtonian viscosity and long-time relaxation phenomena characteristic of glassy polymers.\cite{book:92M, art:02YO}
Hereafter, unless otherwise stated, we measure the physical quantities with units of length $\sigma$, time $\sqrt{m\sigma^2/\epsilon}$, and temperature $\epsilon/k_B$.

In this section, we clarify the linear viscoelastic properties of the model polymer melt.
Figure \ref{fig_geq} shows the stress relaxation function $G(t)$ in the quiescent state.
The stress relaxation function $G(t)$ of the model polymer melt was obtained in Ref. \onlinecite{art:02YO} (although we recomputed $G(t)$ in the present study in order to calculate the accurate linear complex moduli as seen below).
The details of the molecular dynamic simulation can be found in Ref. \onlinecite{art:02YO}.
The stress relaxation function $G(t)$ is calculated as

\begin{equation}\label{eq_gt}
G(t)=<\Pi_{xy}(t+t_0)\Pi_{xy}(t_0)>/k_B T V,
\end{equation}

where $\Pi_{xy}$ is the space integral of the microscopic stress tensor in the volume $V$.
In the figure, the $\alpha$ relaxation time $\tau_\alpha$ and the Rouse relaxation time $\tau_R$ in the quiescent state, where $\tau_\alpha$=310 and $\tau_R$=$6\times 10^4$, are also plotted.
The $\alpha$ relaxation time $\tau_\alpha$ represents the characteristic time of the structural relaxation of bead particles and is calculated from the van Hove self-correlation function, and the Rouse relaxation time $\tau_R$ represents the characteristic time of the conformational relaxation of polymer chains and is calculated from the time-correlation function of the end-to-end vector of each polymer chain.
After the early oscillating behavior for $t \lesssim 10$, which corresponds to the vibrations of the bonds of bead particles on each chain, $G(t)$ exhibits the stretched exponential relaxation of the Kohlrausch-Williams-Watts (KWW) form $G_s(t)$,

\begin{equation}\label{eq_gs}
G_s(t)=G_0\exp[-(t/\tau_s)^c].
\end{equation}

$G(t)$ can be nicely fitted to Eq. (\ref{eq_gs}) with $G_0$=5, $c$=0.5, and $\tau_s=90$ ($\simeq 0.33\tau_\alpha$) for 1$\lesssim t \lesssim 10^3$.
Ultimately, $G(t)$ follows the Rouse dynamics characterized by the Rouse relaxation time $\tau_R$.
The Rouse relaxation function $G_R(t)$ is written as

\begin{equation}\label{eq_gr}
G_R(t)=\frac{\rho_0 T}{N_{\rm b}}\sum_{p=1}^{N_{\rm b}-1}\exp(-2t/\tau_p),
\end{equation}

where $\tau_p/\tau_R=\sin^2(\pi/2N_{\rm b})/\sin^2(\pi p/2N_{\rm b})$ for $p=1,\cdots,N_{\rm b}$.
Thus, in the quiescent state, the stress relaxation function $G(t)$ can be approximately described by the sum of the stretched relaxation function $G_s$ in Eq. (\ref{eq_gs}) and the Rouse relaxation function $G_R(t)$ in Eq. (\ref{eq_gr}), i.e., $G(t)\simeq G_s(t) + G_R(t)$, except the period of initial oscillating modes, and decreases so rapidly that it becomes negligible in the late stage, while $G_R(t)$ is so small as to be negligible in the early stage but can describe the late stage of $G(t)$.

The frequency-dependent shear moduli, i.e., the storage modulus $G'(\omega)$ and the loss modulus $G''(\omega)$, in the linear response regime are obtained by the Fourier transform of the stress relaxation function $G(t)$ in the quiescent state as

\begin{subequations}\label{eq8}
\begin{align}
G'(\omega)&=\omega\int_0^\infty G(t) \sin\omega t \,dt,
\\
G''(\omega)&=\omega\int_0^\infty G(t) \cos\omega t \,dt.
\end{align}
\end{subequations}

$G'(\omega)$ and $G''(\omega)$ represent the elasticity and viscosity of the melt, respectively.
Figure \ref{fig_gpgpp} shows the linear storage modulus $G'(\omega)$ and loss modulus $G''(\omega)$ versus the frequency $\omega$.
The crossover of $G'(\omega)$ and $G''(\omega)$ is observed at $\omega\simeq\tau_\alpha^{-1}$.
The model polymer melt is rather elastic ($ G' > G''$) for $\omega > \tau_\alpha^{-1}$, while it is rather viscous ($G''>G'$) for $\omega< \tau_\alpha^{-1}$.
At the low frequency, i.e., $\omega \lesssim \tau_R^{-1}$, the loss modulus $G''$ is quite dominant and almost proportional to $\omega$.
This indicates that, as the frequency $\omega$ is smaller than $\tau_R^{-1}$, the dynamic viscosity $\eta'$ defined as $\eta'=G''/\omega$ is approximately constant so as to be a simple viscous fluid with a constant viscosity $\eta'$.
The shear moduli calculated from the sum of the stretched exponential form $G_s(t)$ and the Rouse relaxation function $G_R(t)$ are also plotted in the figure.
It is seen that, in the linear response regime, the moduli of the model polymer melt can be well described by those calculated from the sum of $G_s(t)$ and $G_R(t)$, except for the high frequency regime as $\omega \gg \tau_\alpha^{-1}$, which corresponds to the oscillation mode of $G(t)$ in the early stage (See Fig. \ref{fig_geq}).

\section{The effect of inertia of the fluid}

\subsection{Problem and simulation method}
\begin{figure*}[t]
\includegraphics[scale=1]{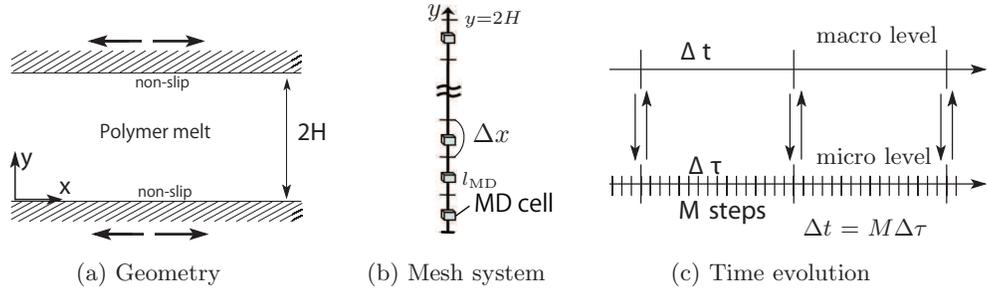}
\caption{
Schematics of the geometry of the problem, the mesh system, and the time-evolution scheme.
}\label{pic_problem}
\end{figure*}

In the rapidly oscillating plates, the oscillatory shear flow becomes non-uniform due to inertia of the fluid via the term $\rho\partial v_x/\partial t$, say the transient force, such that the local rheological property spatially varies according to the local velocity field.
In this section, we investigate the dynamic rheology of the melt in the non-uniform oscillatory shear flow between rapidly oscillating plates (see Fig. \ref{pic_problem}(a)). 
The upper- and lower- plate start to oscillate in the $x$-direction at a time $t$=0 as, respectively,

\begin{equation}\label{eq_vw}
v_w(t)=\mp \Gamma_0 H \omega_0 \cos(\omega_0 t),
\end{equation}

where $\Gamma_0$ is the amplitude of the strain of the system and $H$ is the half of width between the upper- and lower- plate.
We assume that the macroscopic quantities are uniform in the $x$- and $z$-directions, $\partial/\partial x$=$\partial/\partial z$=0.
The macroscopic velocity $v_\alpha$ is described by the following equations,

\begin{equation}\label{eq_mac}
\rho_0\frac{\partial v_x}{\partial t} = 
\frac{\partial \sigma_{xy}}{\partial y},
\end{equation}

and $v_y$=$v_z$=0, where $t$ is the time and $\sigma_{xy}$ is the shear stress. 
We also assume a non-slip boundary condition on each plate.
In the present study, we fix the width between the plates at $2H$=5000.
For the strain amplitude of the system $G_0$, $\Gamma_0$=0.5 is the value that is mainly utilized, while $G_0$=0.02 is a subsidiary used for the comparison.
However, the oscillation frequency of the plate $\omega_0$ is
widely varied in order to investigate the effect of the changing oscillation
frequency on the rheological property of the melt.
The oscillation frequencies $\omega_0$ used in the present simulations are shown by diamonds $\Diamond$ on the upper-axis around $\omega_0\sim\tau_\alpha^{-1}$ in Fig. \ref{fig_gpgpp}.

We solve this problem by using a hybrid simulation of the molecular dynamics (MD) and computational fluid dynamics (CFD).
The details of the method can be found in Ref. \onlinecite{art:10YY}.
Here, we briefly explain the hybrid simulation method.
We calculate the macroscopic velocity in Eq. (\ref{eq_mac}) by using a usual finite volume scheme with a uniform mesh system (see Fig. \ref{pic_problem}(b) ).
However, instead of using any constitutive equation for $\sigma_{xy}$, we calculate the local stresses in small MD cells associated with each mesh interval according to the local strain rates, which are calculated at the CFD level, at each time step of the CFD simulation.\cite{book:89AT, book:08EM}
The MD simulations are performed in each MD cell for the duration of the time-step size of the CFD calculation, and the molecular configurations obtained in each MD cell after one MD run are memorized as the initial configurations of molecules for the next MD run in each MD cell
(see Fig. \ref{pic_problem}(c) ).
By using this method, one can reproduce the memory effect caused by the slow dynamics of the molecular conformation correctly.
In the present study, we divide the lower half between the plates into 128 mesh intervals with a mesh size of $\Delta x$=19.5 and use the symmetric condition at the middle between the plates for the CFD calculation.
For the MD simulation, we use a cubic MD cell with a side length $l_{\rm MD}$=10 so that each MD cell contains 1000 bead particles.
The ratio of the mesh size of CFD $\Delta x$ to the side length of the MD cell $l_{\rm MD}$, which represents the efficiency of the hybrid simulation compared to the full MD simulation, is $\Delta x/l_{\rm MD}$=1.95.
The time-step size of the CFD calculation $\Delta t$ and that of the MD calculation$\Delta \tau$ are fixed at $\Delta t$=1 and $\Delta \tau$=0.001, respectively. Thus, one thousand time steps are carried out in each MD run at each time step of the CFD simulation ($M=1000$ in Fig. \ref{pic_problem} (c)).

\subsection{Velocity profile}
\begin{figure}
\includegraphics[scale=1]{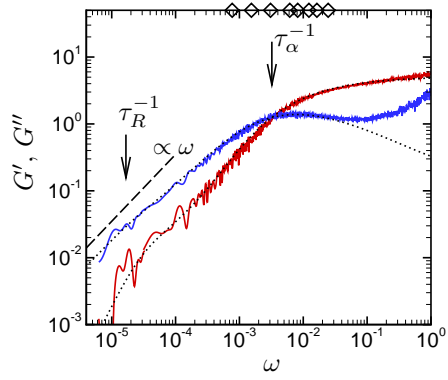}
\caption{(Color)
The storage and loss moduli, $G'(\omega)$ (red solid line) and $G''(\omega)$ (blue solid line), respectively, in the linear response regime.
The dashed line shows the slope of 1 as a guide.
The dotted lines show the shear moduli calculated from the superposition of the stretched exponential form $G_s(t)$ and the Rouse relaxation function $G_R(t)$, $G_s(t)+G_R(t)$.
The diamonds aligned on the upper horizontal axis indicate the values of the oscillation frequencies to be used in the hybrid simulations.
}\label{fig_gpgpp}
\end{figure}

Figure \ref{fig_vx} (a) shows snapshots of the velocity profile of the melt for $\omega_0$=0.0015 and $\Gamma_0$=0.5.
Due to the transient force, the amplitude of the oscillatory velocity rapidly decreases with distance from the oscillating plate, and a thin boundary layer forms over the oscillating plate.
The thickness of the boundary layer $l_b$, which is defined as $v_x(y=l_b)/v_0=e^{-1}$, is also shown at different oscillation frequencies $\omega_0$ for $\Gamma_0$=0.02 and 0.5 in Fig. \ref{fig_vx} (b).
The solid line shows the thickness of the boundary layer for the linear viscoelastic fluid with a linear storage and loss modulus shown in Fig. \ref{fig_gpgpp}.
As is seen in Fig. \ref{fig_gpgpp}, at the low frequency (i.e., $\omega \lesssim \tau_R^{-1}$), the storage modulus $G'$ is much smaller than the loss modulus $G''$ and the dynamic viscosity $\eta'$, which is calculated as $G''/\omega$, becomes almost constant.
Thus, the slope of $\omega^{-1/2}$ for the Newtonian fluid arises at the low frequency in Fig. \ref{fig_vx} (b).
The thickness of the boundary layer of the present polymer melt is close to that of the linear viscoelastic fluid at low oscillation frequencies for the small strain amplitude of the system, $\Gamma_0=0.02$; however, as the oscillation frequency increases, the boundary layer of the melt becomes much thinner than that of the viscoelastic fluid.
The thickness of boundary layer for the large strain amplitude,
$\Gamma_0=0.5$, is thinner than that for the small strain amplitude,
$\Gamma_0=0.02$ at any of the oscillation frequencies.
These features are caused by the shear thinning of the melt because the local strain becomes larger near the oscillating plate as the oscillation frequency $\omega_0$ and
the strain amplitude of the system $\Gamma_0$ increase.
Due to the emergence of the boundary layer, the rheological properties of the melt become more complex in nature, as we see below.

\subsection{Storage and loss modulus}
In this subsection, we investigate the ``local'' rheological properties of the melt in the slab in terms of the storage modulus $G'$ and loss modulus $G''$.
The local moduli are calculated from the first harmonics of the Fourier series of the time evolutions of the local shear stress and strain.
By using the Fourier transform of the time evolution of the local strain $\gamma(y,t)$ and selecting the mode of the oscillation frequency of the plate $\omega_0$, we can approximate the time evolution of the local strain $\gamma$ in the form of $\gamma=\gamma_0(y)\cos(\omega_0t+\psi(y))$.
Here, $\gamma_0(y)$ is the strain amplitude, and $\psi(y)$ is the phase retardation.
In the same way, the local shear stress can also be written as
$\sigma_{xy}=\sigma'(y)\cos(\omega_0t+\psi)-\sigma''\sin(\omega_0t+\psi)$.
The local storage modulus and loss modulus are obtained as
$G'(y)=\sigma'(y)/\gamma_0(y)$ and $G''(y)=\sigma''(y)/\gamma_0(y)$, respectively.
We note that the contribution of the higher harmonics is also important
in the large amplitude oscillatory shear (LAOS) regime and is actually 
detected in the present simulations.
However, the amplitude of the higher harmonics is smaller than that of the first harmonics; the fractional power spectrum of the higher harmonics of the local shear stress is at most 2.8\% in the present simulations.
Thus, the storage and loss modulus represent the basic viscoelastic properties of the melt in the slab.
(The non-linear effect of higher harmonics is investigated in the next subsection.)
In the present subsection, we fix the strain amplitude of the system as $\Gamma_0=0.5$ and change the oscillation frequency of the plate $\omega_0$ variously in order to investigate the dependency of the local rheological property on the oscillation frequency of the plate.

Figure \ref{fig_gpgpp_gmgmd_omw} shows the dependency of the storage and loss modulus on the oscillation frequency, $G'$ and $G''$ vs. $\omega_0$ and that of the strain and strain rate amplitudes on the oscillation frequency, $\gamma_0$ and $\dot\gamma_0$ vs. $\omega_0$, at a fixed position. 
Figure \ref{fig_gpgpp_gmgmd_omw} (a) shows the storage and loss modulus versus oscillation frequency and the strain and strain rate amplitudes versus oscillation frequency at far distances from the plate and (b) those at the near distances from the plate.
It can be seen that, at the high oscillation frequency, say $\omega_0=0.01$, the storage modulus $G'$ is larger than the loss modulus $G''$ at large distances from the plate, e.g., $y=1309$ (Blue) and 771 (Green) in Fig. \ref{fig_gpgpp_gmgmd_omw} (a), while $G'$ is smaller than $G''$ as the distance decreases, e.g., $y=381$ (Red) in Fig. \ref{fig_gpgpp_gmgmd_omw} (a) and at smaller distances in Fig. \ref{fig_gpgpp_gmgmd_omw} (b).
In close vicinity to the plate, e.g., $y=10$ in Fig. \ref{fig_gpgpp_gmgmd_omw} (b), the viscosity is quite dominant, $G''\gg G'$.
Thus, the local rheological properties of the melt vary considerably according to the local flow fields such that three different rheological regimes, i.e., the viscous fluid regime, the liquid-like viscoelastic regime, and the solid-like viscoelastic regime, are formed over the rapidly oscillating plate.
On the contrary, at a low oscillation frequency, say $\omega_0=0.001$, the differences of the storage modulus $G'$ and loss modulus $G''$ are not as large as those at a high oscillation frequency at any distance, and the loss modulus $G''$ is larger than the storage modulus $G'$. 
The differences of local modulus between the different distances are also not as large as those at a high oscillation frequency.
Thus, the local rheological properties of the melt vary moderately between the plates, and the melt has liquid-like viscoelastic behaviors throughout the slab.

At large distances from the plate [Fig. \ref{fig_gpgpp_gmgmd_omw} (a)], the local strain $\gamma_0$ monotonically decreases as the oscillation frequency $\omega_0$ increases because, as we have seen in Fig. \ref{fig_vx}, a thin boundary layer forms over the rapidly oscillating plate, and the thickness of the boundary layer becomes thinner as the oscillation frequency $\omega_0$ increases.
Thus, the local moduli deviate from the linear moduli more at the lower oscillation frequencies due to the shear thinning, while they are rather close to the linear values at the high oscillation frequencies.

The shear thinning behaviors of the local moduli to the local strain amplitude $\gamma_0$ are shown in Fig. \ref{fig_gpep_gm}.
It is seen that the storage modulus $G'$ decreases when the local strain $\gamma_0$ is larger than about 2\%, while the loss modulus $G''$ starts to decrease at a larger strain amplitude ($\gamma_0>2\%$).
The non-monotonic behavior of the loss modulus $G''$ on the strain amplitude
$\gamma_0$ is also observed at a high oscillation frequency [Fig. \ref{fig_gpep_gm} (c)]; the weak shear thickening occurs at a small strain amplitude $\gamma_0\lesssim 0.1$, and then, the shear thinning occurs at a large strain amplitude $\gamma_0>0.1$.
The storage modulus $G'$ decreases more rapidly than the loss modulus $G''$.
These features of the shear thinning behaviors can explain the crossover behavior of the local storage modulus and loss modulus shown in Fig. \ref{fig_gpgpp_gmgmd_omw} (a).
At the high oscillation frequencies, both local moduli are close to the linear moduli because the local strain amplitude is smaller than about 2 \%.
In the linear regime, the storage modulus $G'$ is larger than the loss modulus $G''$ at high oscillation frequencies.
The local storage modulus $G'$ deviates from the linear modulus as the oscillation frequency $\omega_0$ decreases, and the local strain $\gamma_0$ exceeds about 2\%, while the local loss modulus $G''$ remains close to the linear modulus at $\gamma_0\sim 2\%$.
Thus, the storage modulus $G'$ crosses over the loss modulus $G''$ at a certain oscillation frequency, say a crossover frequency $\omega_0^c$.
The loss modulus $G''$ also starts to decrease as the oscillation frequency $\omega_0$ is smaller than the crossover frequency $\omega_0^c$.
However, because the storage modulus $G'$ decreases more rapidly than the loss modulus $G''$ as the strain amplitude $\gamma_0$ increases, the storage modulus $G'$ is smaller than the loss modulus $G''$ for oscillation frequencies smaller than the crossover frequency, $\omega_0<\omega_0^c$.

Figure \ref{fig_deb_omw} shows the local Deborah numbers, $De^\alpha$ and $De^R$, versus the oscillation frequency $\omega_0$.
The local Deborah numbers, $De^\alpha$ and $De^R$, are defined by the products of the oscillation frequency $\omega_0$ and the shear-dependent $\alpha$ relaxation time $\tau_\alpha(\dot\gamma_0)$, $De^\alpha=\omega_0\tau_\alpha(\dot\gamma_0)$, and the shear-dependent Rouse relaxation time $\tau_R(\dot\gamma_0)$, $De^R=\omega_0\tau_R(\dot\gamma_0)$, respectively.
Here, we use the fitting functions for the relaxation times $\tau_\alpha$ and $\tau_R$ for the simple shear flows obtained in Ref. \onlinecite{art:02YO}.
It is seen that the local Deborah numbers $De^\alpha$ are equal to unity around the crossover frequencies for each local moduli $\omega_0^c$.
In the lower figure in Fig. \ref{fig_gpgpp_gmgmd_omw} (a), the local strain rate $\dot\gamma_0$ increases as the distance from the plate $y$ decreases, while it does not so much depend on the oscillation frequency $\omega_0$ but only slightly decreases as the oscillation frequency $\omega_0$ increases.
The $\alpha$ relaxation time $\tau_\alpha(\dot\gamma)$ is the monotonically decreasing function on the strain rate $\dot\gamma$.
Hence, in Fig. \ref{fig_deb_omw} (a), the local Deborah number $De^\alpha$ decreases as the distance from the plate $y$ decreases but does not alter the shape of the curve very much, such that the oscillation frequency
at which the local Deborah number $De^\alpha$ is equal to unity, i.e., the crossover frequency $\omega_0^c$, shifts to a higher value as the distance $y$ decreases.
As the local Deborah number $De^\alpha$ is less than unity, the loss modulus $G''$ also decreases as the oscillation frequency $\omega_0$ decreases as is seen in Fig. \ref{fig_gpgpp} for the linear moduli.

The behaviors of the storage modulus and loss modulus and the amplitude of the local strain and strain rate versus oscillation frequency near the plate are shown in Fig. \ref{fig_gpgpp_gmgmd_omw} (b).
The dependency of the local strain and strain rate, $\gamma_0$ and $\dot\gamma_0$, on the oscillation frequency $\omega_0$ near the plate is quite different from those far from the plate.
The local strains $\gamma_0$ at $y$=88 and 185 slightly increase with the oscillation frequency in the low oscillation frequencies, e.g., $\omega_0\lesssim 0.002$, but decrease as the oscillation frequency increases in $\omega_0\gtrsim 0.002$, while the local strain at $y$=10 monotonically increases with the oscillation frequency.
The local strain rates $\dot\gamma_0$ monotonically increases with the oscillation frequency in the close vicinity of the plate, i.e, $y$=88 and 10, while the strain rate at $y$=185 shows the non-monotonic dependency on the oscillation frequency.
The spatial variations of local strains and local strain rates are small at low oscillation frequencies and are large at high oscillation frequencies.
This feature also holds for the local moduli.

In Fig. \ref{fig_deb_omw} (b), we show the local Deborah numbers near the plate.
The non-monotonic dependency of the local storage modulus $G'$ on the oscillation frequency $\omega_0$, which can be seen in Fig. \ref{fig_gpgpp_gmgmd_omw} (b), e.g., $G'$ at $y$=88 and 185, might be related to the local Deborah number $De^R$ in Fig. \ref{fig_deb_omw} (b).
As can be seen in Fig. \ref{fig_gpgpp_gmgmd_omw} (b), the local strain rate
$\dot\gamma_0$ rapidly increases as the oscillation frequency $\omega_0$ increases at low oscillation frequencies; however, at high oscillation frequencies, it shows a different dependency on the oscillation frequency $\omega_0$. The local strain rate slightly decreases at $y$=185, does not change much at $y$=88, and monotonically increases at $y$=10, as the oscillation frequency $\omega_0$ increases.
The Rouse relaxation time $\tau_R$ monotonically decreases as the strain rate $\dot\gamma_0$ increases.
Hence, the local Deborah number $De^R$, $De^R=\omega_0\tau_R(\dot\gamma_0)$, decreases even if the oscillation frequency $\omega_0$ increases at low oscillation frequencies because the Rouse relaxation time $\tau_R$ rapidly decreases.
On the contrary, at high oscillation frequencies, $De^R$ increases at $y$=88 and 185 as the oscillation frequency $\omega_0$ increases because $\tau_R$ dose not change much or rather increases slightly.
The elasticity grows as the local Deborah number $De^R$, but it becomes negligible as $De^R$ is less than unity.
Thus, the local storage modulus $G'$ at $y$=88 and 185 in Fig. \ref{fig_gpgpp_gmgmd_omw} (b) varies according to the local Deborah number $De^R$, but the local storage modulus $G'$ at $y$=10 becomes negligibly small except at low oscillation frequencies.

In Fig. \ref{fig_etap_gmd}, we also show shear thinning behavior of the local dynamic viscosity $\eta'$(=$G''/\omega_0$) to the local strain rate $\dot\gamma_0$ for various oscillation frequencies, i.e., $\omega_0$=$1.5\times 10^{-3}$, $6.1\times 10^{-3}$, $1.2\times 10^{-2}$, and $2.5\times 10^{-2}$.
It is seen that, at large strain rates, e.g., $\dot\gamma_0 \gtrsim 0.01$, the dependence of the dynamic viscosity $\eta'$ on the oscillation frequency $\omega_0$ is weakened due to which the results for different oscillation frequencies coincide with each other for $\dot \gamma_0\gtrsim$ 0.1.
A second Newtonian regime is also observed at very large strain rates, e.g., $\dot\gamma_0 > 0.4$.
We also find that the slope of the shear thinning is similar to that observed in steady shear flows, in which the slope is about -0.7\cite{art:02YO}.
Thus, the shear thinning behavior similar to that of the steady shear flows is observed in the vicinity of the plate at high oscillation frequencies.

We also show a diagram of the loss tangent $\tan\delta$, which is defined as the ratio of the viscosity to the elasticity, $\tan\delta=G''/G'$, for different oscillation frequencies and local strain rates in Fig. \ref{fig_losstan_omw_gmd}.
In the diagram, the dashed and dotted lines show that the inverses of the shear-dependent $\alpha$ and Rouse relaxation times, $\tau_\alpha(\dot\gamma)^{-1}$ and $\tau_R(\dot\gamma)^{-1}$, equal the oscillation frequency $\omega_0$; i.e., on the dashed and dotted lines, the local Deborah numbers $De^\alpha$ and $De^R$ are equal to unity, respectively.
The upper side of the diagram indicates the smaller distance from the plate while the lower side the larger distance because the local strain rate decreases as the distance from the plate increases.
No symbols are plotted for large and small strain rates at low oscillation frequencies because no data are available given that the spatial variations of local strain rates are small at low oscillation frequencies as we have seen in Fig. \ref{fig_gpgpp_gmgmd_omw}.
Below the dashed line, the loss tangent is less than unity, while above the dashed line, the loss tangent is larger than unity.
Thus, the crossover of the storage modulus and loss modulus takes place at the dashed line.
Near or above the dotted line, the loss tangent is quite large, and the elasticity may be negligible.
Hence, the melt behaves as a viscous fluid provided that the value of the local strain rate is larger than or close to values that lie on the dotted line.
Thus, the melt forms three different rheological regimes, i.e., the solid-like viscoelastic, liquid-like viscoelastic, and viscous fluid regimes, according to the local strain rates and oscillation frequencies.

\subsection{LAOS analysis}
As the local strain increases near the oscillating plate, the amplitude of higher harmonics of the local macroscopic quantities becomes large, and thus, the time evolution of the local quantities deforms evidently from that of pure sinusoidal curves.
In the present subsection, we carry out LAOS analysis\cite{art:02HKAL,art:05CHAL,art:08EHM} to examine the non-linear effects of higher harmonics on the local macroscopic quantities.

Figure \ref{fig_powspec} shows the power spectra of the local shear stress and strain rate, $|\tilde \sigma_{xy}(\omega)|^2$ and $|\tilde{\dot\gamma}(\omega)|^2$, in the rapidly oscillating plates with an oscillation frequency $\omega_0=0.025$.
Here, $\tilde \sigma_{xy}(\omega)$ and $\tilde {\dot\gamma}(\omega)$ represent the Fourier coefficients of the shear stress and strain rates, respectively.
The peaks of the higher harmonics are detected in the odd harmonics, $3\omega_0$, $5\omega_0$, $\cdots$, near the oscillating plate but disappear far from the plate.
The higher harmonics arise both in the local stress and strain rate because the higher harmonics of the local strain rate is induced by the local stress with the higher harmonic contribution and the higher harmonics of the local stress is also induced by that local strain rate.

Figure \ref{fig_ylb_powspec} shows the spatial variation of the fractional amplitude of the third harmonics to the basic oscillation for various oscillation frequencies $\omega_0$ at $\Gamma_0=0.5$.
The horizontal axis is the normalized distance with respect to the thickness of boundary layer shown in Fig. \ref{fig_vx}.
Even though the local strain monotonically increases while approaching the oscillating plate, the fractional amplitude of the third harmonics rather decreases rapidly in the boundary layer, i.e., $y\lesssim l_b$.
This is caused by the boundary effect because the velocity at the oscillating plate is purely sinusoidal, and higher harmonics are not allowed in the local strain rate on the oscillating plate.
Thus, the contribution of the higher harmonics for the local stress is depressed in the boundary layer.
The fractional amplitude of the third harmonics takes the maximum value at the outside of the boundary layer for each oscillation frequency, $y>l_b$, and then decreases as the normalized distance increases.
We also show the fractional amplitude of the third harmonics versus local strain amplitude at different strain amplitudes of the system, $\Gamma_0$=0.5 and 0.02, for various oscillation frequencies $\omega_0$ in Fig. \ref{fig_gm_powspec}. 
For a large strain amplitude of the system, $\Gamma_0=0.5$, the fractional amplitude of the third harmonics has a maximum value around the position where the local strain amplitude is unity, $\gamma_0\simeq 1$, and rapidly decreases in the thickness of boundary layer, $y\lesssim l_b$.
On the contrary, for a small strain amplitude of the system $\Gamma_0=0.02$, the maximum occurs not at $\gamma_0\simeq 1$ but around the position that the value of the distance from the plate coincides with the value of the thickness of boundary layer, $y\simeq l_b$.
For $\Gamma_0=0.02$, the position at which the local strain amplitude is unity lies at the inside of the boundary layer.
Thus, the intrinsic maximum of the fractional amplitude of the third harmonics for $\gamma_0\simeq 1$ is suppressed inside the thickness of boundary layer, $y \lesssim l_b$.

Figure \ref{fig_gm_pxy} shows the Lissajous-Bowditch curves of local shear stress $\sigma_{xy}$ versus local strain $\gamma$ at different oscillation frequencies of the plate $\omega_0$ with $\Gamma_0=0.5$.
The dotted line (green) indicates a perfect ellipse drawn by the storage and loss modulus calculated from the Fourier coefficients of the first harmonics.
Thus, the deviation of the Lissajous-Bowditch curve from the pure ellipse represents the contribution of the higher harmonics in each figure.
At each oscillation frequency, the deviation is more evident at some distance away from the plate than in close vicinity to the plate.
This agrees with the fact that the contribution of the higher harmonics is depressed inside the boundary layer and that the fractional amplitude of the third harmonics assumes a maximum at the outside of boundary layer.
In Fig. \ref{fig_gm_pxy}, the thicknesses of the boundary layer $l_b$ are $l_b$=223, 45, and 20 for $\omega_0$=$1.5\times 10^{-3}$, $6.1\times 10^{-3}$, and 0.025, respectively.
We also show the minimum-strain modulus $G'_M$ and large-strain modulus $G'_L$ defined as $G'_M=d \sigma_{xy}/d \gamma |_{\gamma=0}$ and $G'_L=\sigma_{xy}/\gamma|_{\gamma=\gamma_0}$,\cite{art:08EHM} respectively.
The minimum-strain modulus $G'_M$ yields the elasticity at the point where the change of strain rate is zero, $d\dot \gamma/dt=0$, and the large-strain modulus $G'_L$ yields the elasticity at the point where the instantaneous strain rate is zero, $\dot \gamma=0$.
Both the minimum-strain $G'_M$ and large-strain moduli $G'_L$ coincide with the elastic modulus $G'$ in the linear regime, $G'_M, G'_L \rightarrow G'$ for small $\gamma_0$.
It is seen that the large-strain modulus $G'_L$ is larger than the minimum-strain modulus $G'_M$ within a cycle of Lissajous-Bowditch curves at the close vicinity of the plate, e.g., $y=10$.
Thus, intra-cycle stiffening occurs in close vicinity of the plate, although the storage modulus $G'$ exhibits shear-thinning behavior for the local strain amplitude $\gamma_0$, i.e., inter-cycle softening (see also Fig. \ref{fig_gpep_gm}).

Finally, we show the 3D Lissajous-Bowditch curve of the local shear stress $\sigma_{xy}$, local strain $\gamma$, and local strain rate $\dot\gamma$ near the rapidly oscillating plate.
The 2D projections on each plane are also shown.
The cycle of $\dot\gamma$-- $\sigma_{xy}$ curve is very narrow, and a secondary loop\cite{art:10EM} is observed at a large instantaneous strain rate.
The $\gamma$--$\dot\gamma$ curve also deviates from conforming to a perfect ellipse because the higher harmonics are also involved in the local strain and strain rate.

\section{summary}
\begin{figure*}[htbp]
\includegraphics[scale=1]{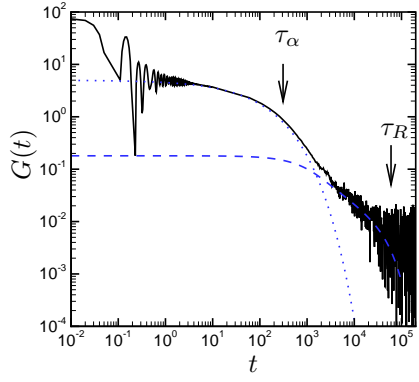}
\caption{(Color online)
The stress relaxation function $G(t)$ of the model polymer melt in the quiescent state (the solid line).
The dotted line shows the stretched exponential form $G_s(t)$ in Eq. (\ref{eq_gs}) with $G_0$=5, $\tau_s$=90, and $c$=0.5.
The dashed line shows that the Rouse relaxation function $G_R(t)$ in Eq. (\ref{eq_gr}) with $\tau_R$=$6\times 10^4$. $\tau_\alpha$ is the $\alpha$ relaxation time, $\tau_\alpha$=310.
}\label{fig_geq}
\end{figure*}

We investigated the dynamic rheology of a model polymer melt in non-uniform oscillatory shear flows under the transient force between rapidly oscillating plates by using a hybrid simulation of the molecular dynamics and computational fluid dynamics.
In the quiescent state, the melt is in a supercooled state, where the stress relaxation function exhibits a stretched exponential form on the time scale of the $\alpha$ relaxation time $\tau_\alpha$ and then follows the Rouse relaxation function characterized by the Rouse relaxation time $\tau_R$. [See Fig. \ref{fig_geq}.]

\begin{figure*}[htbp]
\includegraphics[scale=1]{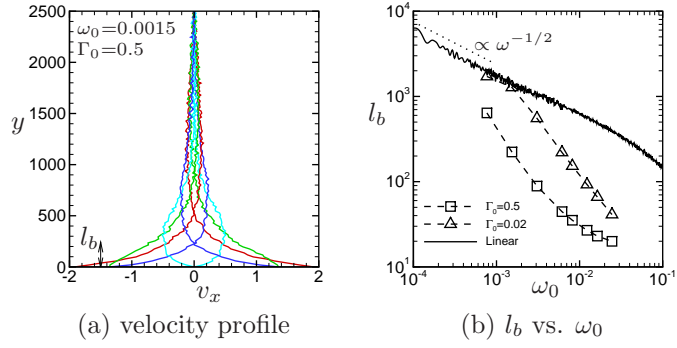}
\caption{(Color online)
The snapshots of the velocity profile for $\omega_0$=0.0015 and $\Gamma_0$=0.5 (a) and the thickness of boundary layer $l_b$ for various oscillation frequencies $\omega_0$ (b).
In figure (b), the solid line, dashed line with squares, and dashed line with triangles show the results for the linear regime ($\Gamma_0 \ll 1$), $\Gamma_0$=0.5, and $\Gamma_0$=0.02, respectively.
The dotted line in figure (b) shows the slope of $\omega^{-1/2}$ for the Newtonian fluid.}\label{fig_vx}
\end{figure*}

In the rapidly oscillating plates, the melt forms a thin boundary layer over the plates due to the transient force [See Fig. \ref{fig_vx}] such that the dynamic rheology of the melt spatially varies considerably according to the local flow field.
At a high oscillation frequency, the melt forms three different rheological regimes, i.e., the viscous fluid regime ($G'\ll G''$), the liquid-like viscoelastic regime ($G'<G''$), and the solid-like viscoelastic regime ($G'>G''$), over the oscillating plates according to the local Deborah number, while, at a low oscillation frequency, the spatial variation of the storage modulus and loss modulus is rather small, and the loss modulus is larger than the storage modulus, $G'<G''$, at any distance from the plate.

\begin{figure}[htbp]
\includegraphics[scale=1]{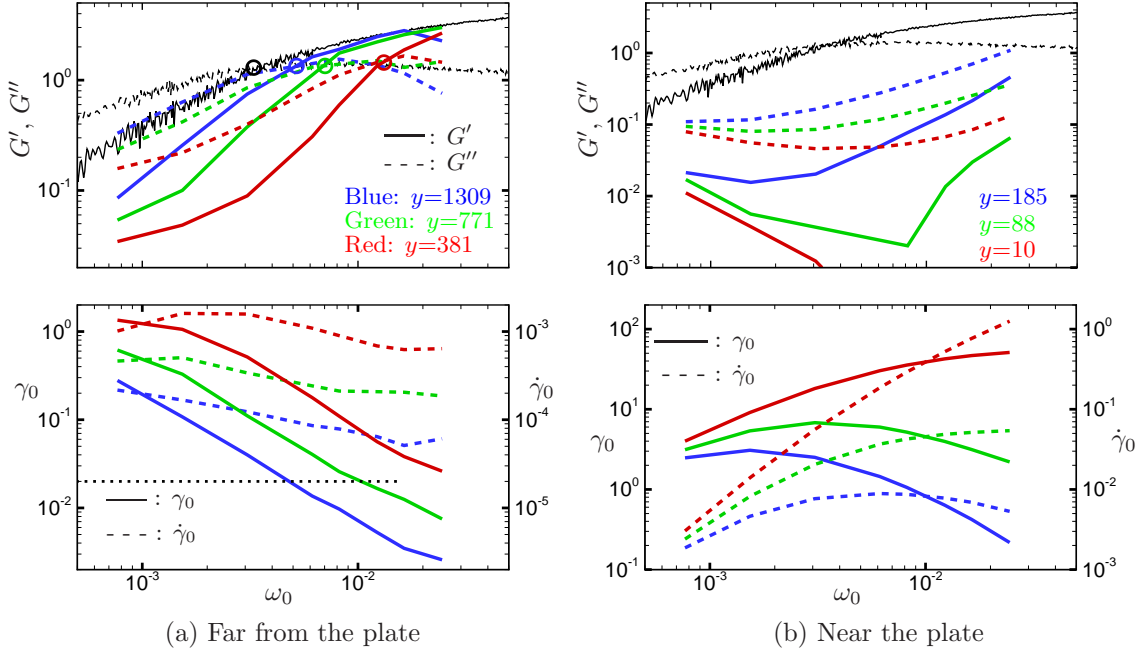}
\caption{(Color)
The local dynamic rheology, $G'$, $G''$ vs. $\omega_0$, and the amplitude of local strain and that of the local strain rate, $\gamma_0$ and $\dot \gamma_0$, respectively, versus the oscillation frequency $\omega_0$ far from the plate (a), e.g., $y$=1309 (Blue), $y$=771 (Green) and 381 (Red), and near the plate (b), e.g., $y=185$ (Blue), $y$=88 (Green) and 10 (Red).
The black lines show the linear moduli depicted in Fig. \ref{fig_gpgpp}.
The open circles in the upper figure of (a) show the crossover points of $G'$ and $G''$.
The dotted lines in the lower figure of (a) show $\gamma_0$=0.02.
}\label{fig_gpgpp_gmgmd_omw}
\end{figure}

\begin{figure}[htbp]
\includegraphics[scale=1]{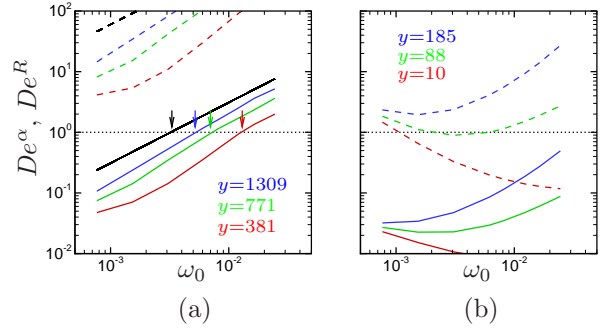}
\caption{(Color)
The local Deborah numbers defined as $De^\alpha$=$\omega_0\tau_\alpha(\dot\gamma_0)$ and $De^R$=$\omega_0\tau_R(\dot\gamma_0)$ far from the plate (a) and near the plate (b).
Here, $\tau_{\alpha,\,R}(\dot\gamma_0)$ are the $\alpha$ and Rouse relaxation times depending on the local strain rate $\dot\gamma_0$.
We use the formulas for $\tau_{\alpha,\,R}(\dot\gamma)$ that are obtained by the molecular dynamic simulations of the model polymer melt in steady shear flows\cite{art:02YO}.
The solid lines show $De^\alpha$ and the dashed lines $De^R$.
The black lines show the Deborah numbers defined via the oscillation
 frequency $\omega_0$ and the relaxation times of the melt in the
 quiescent state $\tau_{\alpha,\,\tau_R}(0)$.
The downward arrows show the crossover frequencies $\omega_0^c$ at each
 distance from the plate, which are also shown by open circles in
 Fig. \ref{fig_gpgpp_gmgmd_omw} (a).
}\label{fig_deb_omw}
\end{figure}

\begin{figure}[htbp]
\includegraphics[scale=1]{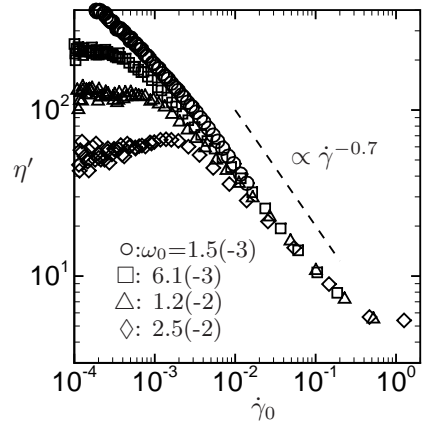}
\caption{
The dynamic viscosity $\eta'$ vs. the amplitude of the strain rate $\dot\gamma_0$ for $\omega_0$=$3.1\times 10^{-3}$, $1.2\times 10^{-2}$, and $2.5\times 10^{-2}$.
Here, $\eta'=G''/\omega$ and $\dot\gamma$=$\omega_0 \gamma_0$.
}\label{fig_etap_gmd}
\end{figure}

The dependency of the local moduli on the oscillation frequency, $G'$ and $G''$ vs. $\omega_0$ at a fixed position changes according to the distance from the plate.
Far from the plate [Fig. \ref{fig_gpgpp_gmgmd_omw} (a)], the local strain $\gamma_0$ decreases as the oscillation frequency $\omega_0$ increases such that the local moduli deviate from the linear moduli larger at a low frequency than at a high oscillation frequency.
Near the plate [Fig. \ref{fig_gpgpp_gmgmd_omw} (b)], the dependency of the local strain and strain rate on the oscillation frequency is quite different from that far from the plate, e.g., in close vicinity to the plate, the local strain and strain rate monotonically increase as the oscillation frequency.
As the local strain rate becomes larger than about 0.01 near the plate, the shear thinning of the dynamic viscosity is increased, and the shear thinning behavior becomes similar to that observed in steady shear flows [see Fig. \ref{fig_etap_gmd}.]

\begin{figure}[htbp]
\includegraphics[scale=1]{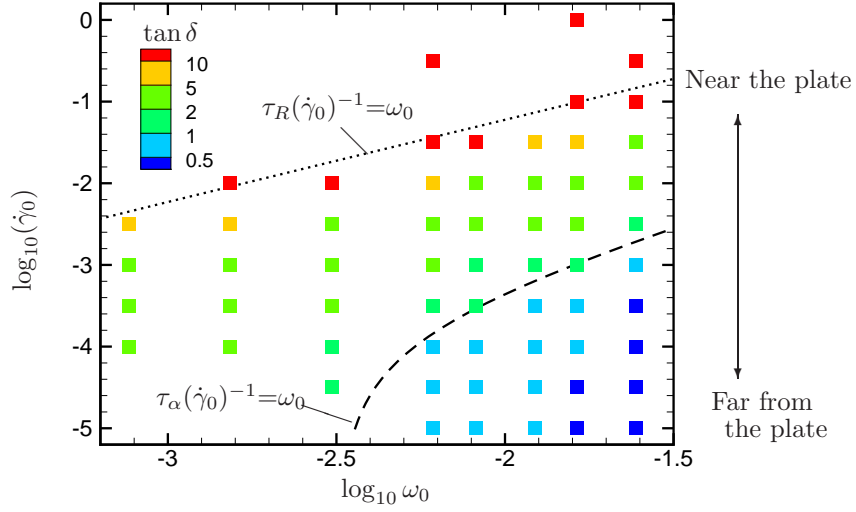}
\caption{(Color)
The amplitude of the loss tangent $\tan\delta$, which is defined as $\tan\delta=\bar G''/\bar G'$, for different oscillation frequencies and local strain rates.
Here, $\bar G'$ and $\bar G''$ are averages of the local storage modulus and loss modulus provided that the local strain rate is in each interval with 0.5 logarithmic scale centered at the value of the local strain rate of each plot, respectively.
On the dashed and dotted lines, the inverses of the shear-dependent $\alpha$ and Rouse relaxation times, $\tau_\alpha(\dot\gamma)$ and $\tau_R(\dot\gamma)$, equal the oscillation frequency, respectively.
}\label{fig_losstan_omw_gmd}
\end{figure}

The diagram of the loss tangent of the melt for different oscillation frequencies and local strain rates is also shown in Fig. \ref{fig_losstan_omw_gmd}.
It is seen in the diagram that the melt generates different rheological regimes according to the oscillation frequency and local strain rates.

\begin{figure}[htbp]
\includegraphics[scale=1]{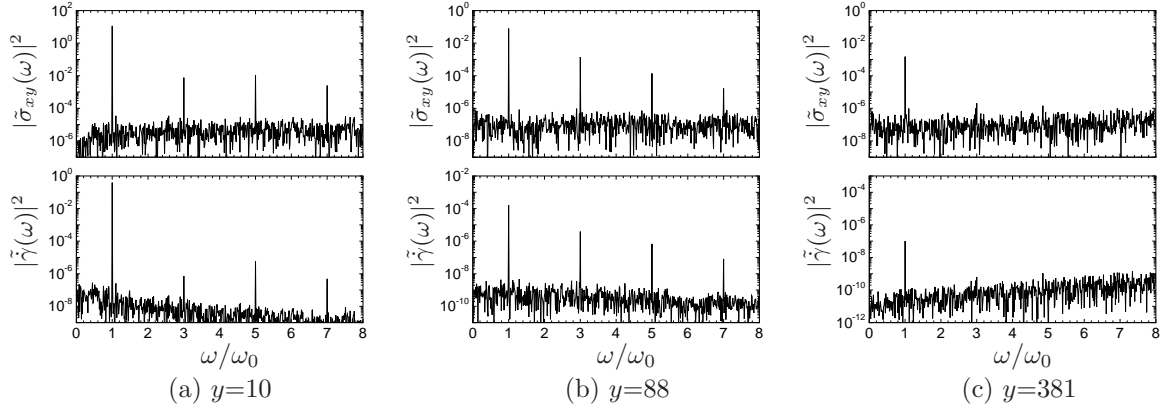}
\caption{
The power spectra of the local shear stress and strain rate, $|\tilde
 \sigma_{xy}(\omega)|^2$ (the upper figures) and $|\tilde {\dot
 \gamma}(\omega)|^2$ (the lower figures), at different positions for
 $\omega_0=0.025$ and $\Gamma_0=0.5$. 
}\label{fig_powspec}
\end{figure}

\begin{figure}[htbp]
\includegraphics[scale=1]{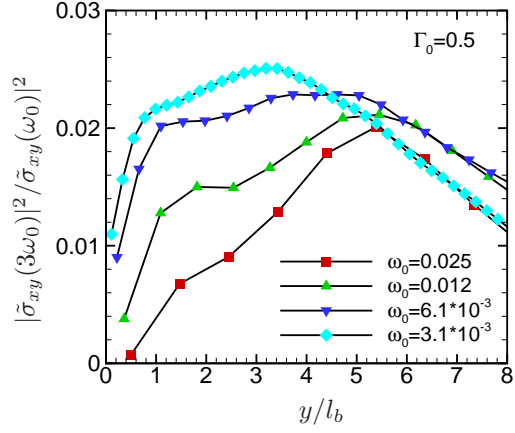}
\caption{(Color online)
The spatial variation of the fractional amplitude of the third
 harmonics to the first harmonics of the local shear stress, $|\tilde
 \sigma_{xy}(3\omega_0)|^2/|\tilde \sigma_{xy}(\omega_0)|^2$, for
 various oscillation frequencies of the plate $\omega_0$ at $\Gamma_0=0.5$.
The horizontal axis shows the distance normalized to the thickness of
 the boundary layer at each oscillation frequency $\omega_0$, shown in Fig. \ref{fig_vx}.
}\label{fig_ylb_powspec}
\end{figure}

\begin{figure}[htbp]
\includegraphics[scale=1]{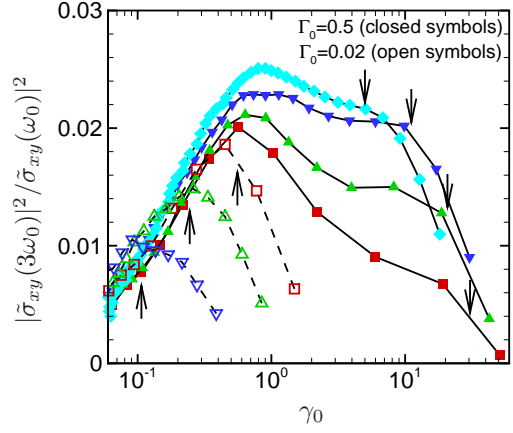}
\caption{(Color online)
The fractional amplitude of the third harmonics to the
 first harmonics of the
 local shear stress, $|\tilde \sigma_{xy}(3\omega_0)|^2/|\tilde \sigma_{xy}(\omega_0)|^2$, versus the
 amplitude of local strain, $\gamma_0$, at the different amplitudes of
 strain of the system $\Gamma_0$=0.5 and 0.02 for various oscillation
 frequencies $\omega_0$.
The shape of the symbol represents the oscillation frequency $\omega_0$, which  is the same as in Fig. \ref{fig_ylb_powspec}.
The downward and upward arrows show the position of $y=l_b$ for each oscillation
 frequency of the plate at $\Gamma_0=0.5$ and $\Gamma_0=0.02$, respectively.
}\label{fig_gm_powspec}
\end{figure}

We also investigate the non-linear rheological properties in the LAOS regime in the vicinity to the oscillating plate.
The odd higher harmonics, $3\omega_0$, $5\omega_0$, $\cdots$, are detected in the power spectra of the local macroscopic quantities near the oscillating plate [see Fig. \ref{fig_powspec}.]
The fractional amplitude of the higher harmonics rapidly decreases inside the boundary layer while approaching the oscillating plate, although the local strain increases monotonically. [See Fig. \ref{fig_ylb_powspec}.]
This is because higher harmonics are not allowed in the velocity on the oscillating plate due to a non-slip boundary, and thus, the higher harmonics of local quantities are suppressed within the boundary layer.

\begin{figure}[htbp]
\includegraphics[scale=1]{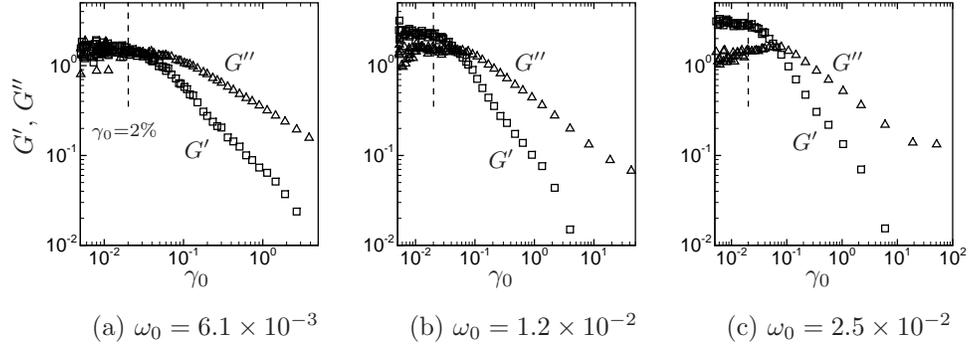}
\caption{
The local storage and loss moduli, $G'$ and $G''$, vs. the local strain
 amplitude $\gamma_0$ for $\omega_0$=$6.1\times 10^{-3}$ (a), $1.2\times
 0^{-2}$ (b), and $2.5\times 10^{-2}$ (c).
}\label{fig_gpep_gm}
\end{figure}

\begin{figure}[htbp]
\includegraphics[scale=1]{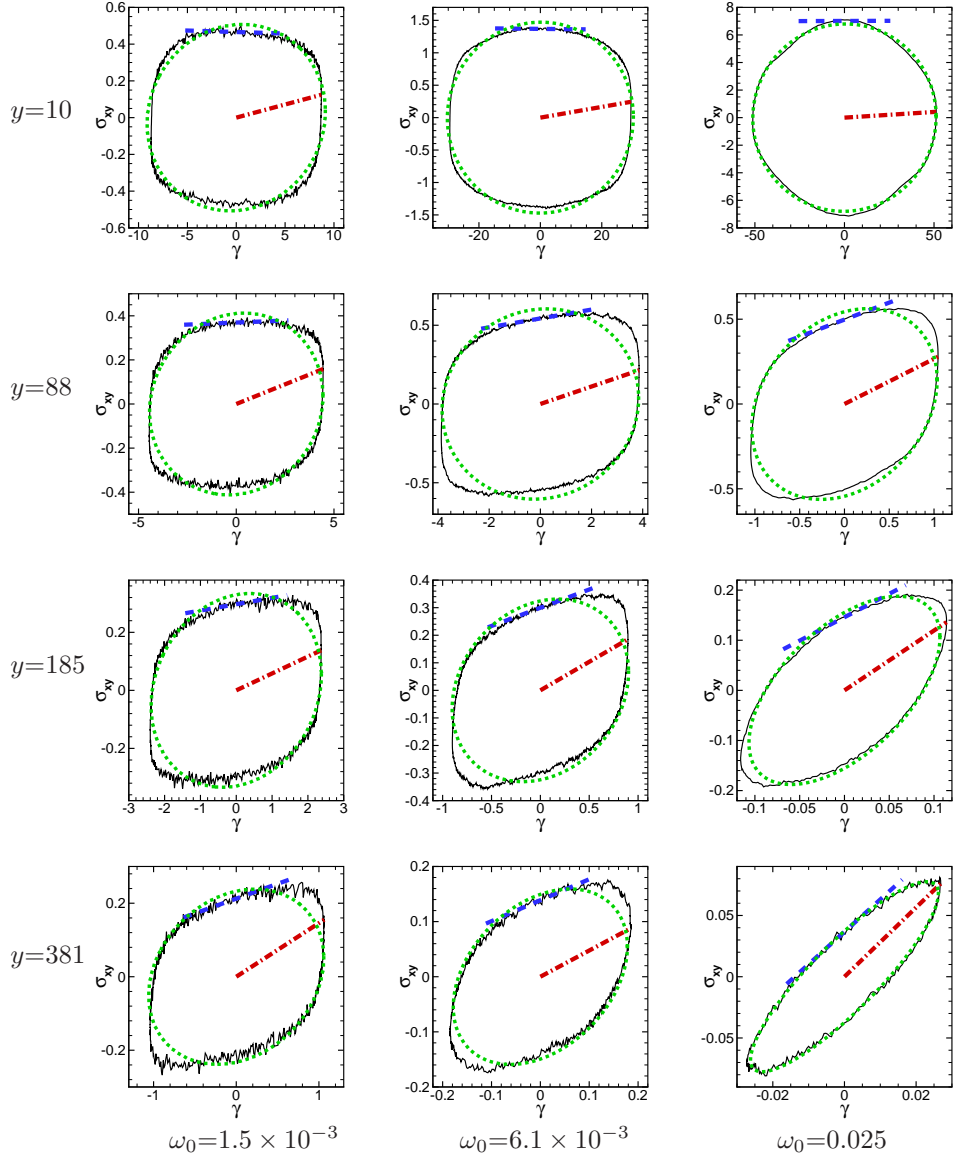}
\caption{(Color)
The Lissajous-Bowditch curves of the local shear stress $\sigma_{xy}$ vs. the local
 strain $\gamma$ at different oscillation frequencies of the plate $\omega_0$ with $\Gamma_0=0.5$.
The dashed line (blue) shows the minimum-strain modulus (or tangent modulus) $G'_M$, $G'_M=d \sigma_{xy}/d\gamma|_{\gamma=0}$, and
the dash-dotted line (red) shows the large-strain modulus (or scant modulus)
 $G'_L$, $G'_L=\sigma_{xy}/\gamma|_{\gamma=\gamma_0}$. 
The dotted curve (green) shows the pure ellipse formed by the storage and
 loss modulus for the first harmonics. 
}\label{fig_gm_pxy}
\end{figure}

The Lissajous-Bowditch curve of the local shear stress versus local strain changes the shape due to the contribution of the higher harmonics depending on the distance from the plate.
The elastic modulus exhibits shear-thinning behavior between the different positions (inter-thinning) [see Fig. \ref{fig_gpep_gm}], but shear-thickening behavior is observed in a cycle of the Lissajous-Bowditch curve at a fixed position in the vicinity of the plate (intra-thickening) [see Fig. \ref{fig_gm_pxy}].

% \appendix

% \section{}
% \label{}
%\bibliographystyle{junsrt}
%\bibliography{art,book}

\begin{thebibliography}{100}
%
\bibitem{book:92M}
S. Matsuoka,
{\it Relaxation phenomena in polymers}
(Oxford, New York, 1992).
%
\bibitem{book:96S}
G. R. Strobl,
{\it The Physics of Polymers}
(Springer, Heidelberg, 1996).
%
\bibitem{book:87BAH}
R. B. Bird, R. C. Armstrong, and O. Hassager,
{\it Dynamics of polymeric liquids} Vol. 1 (John Wiley and Sons, New York, 1987).
%
\bibitem{book:88L}
R. G. Larson,
{\it Constitutive equations for polymer melts and solutions} (Butterworths, Boston, 1988).
%
\bibitem{art:77S}
J. L. Schrag,
``Deviation of velocity gradient profiles from the ``Gap Loading'' and
	``Surface Loading'' Limits in dynamic simple shear
	experiments'',
Trans. Soc. Rheol. {\bf 21}, 399 (1977).
%
\bibitem{art:96D}
J. Dunwoody,
``The effects of inertia and finite amplitude on oscillatory plane shear
	flow of K-BKZ fluids such as LDPE melts'',
J. Non-Newtonian Fluid Mech. {\bf 65}, 195 (1996).
%
\bibitem{art:98YGSD}
J. A. Yosick, J. A. Giacomin, W. E. Stewart, and F. Ding,
``Fluid inertia in large amplitude oscillatory shear'',
Rheol Acta {\bf 37}, 365 (1998).
%
\bibitem{art:99DGBK}
F. Ding, A. J. Giacomin, R. B. Bird, and C. B. Kweon,
``Viscous dissipation with fluid inertia in oscillatory shear flow'',
J. Non-Newtonian Fluid Mech. {\bf 86}, 359 (1999).
%
%\bibitem{art:1867M}
%J. C. Maxwell,
%``On the dynamical theory of gases'',
%Phil. Trans. Roy. Soc. {\bf A157}, 49 (1867).
%%
%\bibitem{book:76J}
%H. Jeffreys,
%{\it The Earth} (Cambridge University Press, New York, 1976).
%%
%\bibitem{art:63WM}
%J. L. White and A. B. Metzner,
%``Development of constitutive equations for polymeric melts and solutions'',
%J. Appl. Polym. Sci. {\bf 7}, 1867 (1963).
%%
%\bibitem{art:82G}
%H. Giesekus,
%``A simple constitutive equation for polymer fluids based on the concept of deformation-dependent tensorial mobility'',
%J. Non-Newtonian Fluid Mech. {\bf 11}, 69 (1982).
%%
\bibitem{art:08YY}
S. Yasuda and R. Yamamoto,
``A model for hybrid simulation of molecular dynamics and computational fluid dynamics'',
Phys. Fluids {\bf 20}, 113101 (2008).
%
\bibitem{art:09YY}
S. Yasuda and R. Yamamoto,
``Rheological properties of polymer melt between rapidly oscillating plates: an application of multiscale modeling'',
EPL {\bf 86}, 18002 (2009).
%
\bibitem{art:10YY}
S. Yasuda and R. Yamamoto,
``Multiscale modeling and simulation for polymer melt flows between parallel plates'',
Phys. Rev. E {\bf 81}, 036308 (2010).
%
\bibitem{art:03EE}
W. E and B. Engquist,
``The heterogeneous multi-scale methods'',
Comm. Math. Sci. {\bf 1}, 87 (2003).
%
\bibitem{art:07EELRV}
W. E, B. Engquist, X. Li, W. Ren and E. Vanden-Eijnden,
``Heterogeneous multiscale methods: a review'',
Commun. Comput. Phys. {\bf 2}, 367 (2007).
%
\bibitem{art:03KGHKRT}
I. G. Kevrekidis, C. W. Gear, J. M. Hyman, P. G. Kevrekidis, O. Runborg, and C. Theodoropoulos,
``Equation-free, coarse-grained multiscale computation: enabling microscopic simulations to perform system-level analysis'',
Comm. Math. Sci. {\bf 1}, 715 (2003).
%
\bibitem{art:09KS}
I. G. Kevrekidis and G. Samaey,
``Equation-free multiscale computation: algorithms and applications'',
Annu. rev. phys. chem. {\bf 60}, 321 (2009).
%
\bibitem{art:05RE}
W. Ren and W. E,
``Heterogeneous multiscale method for the modeling of complex fluids and micro-fluidics'',
J. Compt. Phys. {\bf 204}, 1 (2005).
%
\bibitem{art:06DFSKK}
S. De, J. Fish, M. S. Shephard, P. Keblinski, and S. K. Kumar,
``Multiscale modeling of polymer rheology'',
Phys. Rev. E {\bf 74}, 030801(R) (2006).
%
\bibitem{art:10KOK}
D. A. Kessler, E. S. Oran, and C. R. Kaplan,
``Towards the development of a multiscale, multiphysics method for the simulation of rarefied gas flows'',
J. Fluid Mech. (in print).
%
%\bibitem{art:93LO}
%M. Laso and H. C. \"Ottinger,
%``Calculation of viscoelastic flow using molecular models: the CONNFFESSIT approach'',
%J. Non-Newtonian Fluid Mech. {\bf 47}, 1 (1993).
%
%\bibitem{art:95FLO}
%K. Feigl, M. Laso, and H. C. \"Ottinger,
%``CONNFFESSIT approach for solving a two-dimensional viscoelastic fluid problem'',
%Macromolecules {\bf 28}, 3261 (1995).
%
%\bibitem{art:97LPO}
%M. Laso, M. Picasso, H. C. \"Ottinger,
%``2-D time-dependent viscoelastic flow calculations using CONNFFESSIT'',
%AIChE J. {\bf 43}, 877 (1997).
%
\bibitem{art:90KG}
K. Kremer and G. S. Grest,
``Dynamics of entangled linear polymer melts: A molecular-dynamics simulation'',
J. Chem. Phys. {\bf 92}, 5057 (1990).
%
\bibitem{art:02YO}
R. Yamamoto and A. Onuki,
``Dynamics and rheology of a supercooled polymer melt in shear flow'',
J. Chem. Phys. {\bf 117}, 2359 (2002).
%
\bibitem{book:89AT}
M. P. Allen and D. J. Tildesley,
{\it Computer Simulation of Liquids},
(Oxford University Press, Oxford, 1989).
%
\bibitem{book:08EM}
D. J. Evans and G. Morris,
{\it Statistical mechanics of nonequilibrium liquids},
(Cambridge university press, New York, 2008).
%
%
%\bibitem{art:84BC}
%D. Brown and J. H. R. Clarke,
%``A comparison of constant energy, constant temperature and constant
% pressure ensembles in molecular dynamics simulations of atomic liquids,''
%Mol. Phys. {\bf 51}, 1243 (1984).
%
%\bibitem{art:05SKK}
%S. Sen, S. K. Kumar, and P. Keblinski,
%``Viscoelastic properties of polymer melts from equilibrium molecular dynamics simulations'',
%Macromolecules {\bf 38}, 650 (2005).
%
%\bibitem{book:97HP}
%R. R. Huilgol and N. Phan-Thien,
%{\it Fluid mechanics of viscoelasticity} (Elsevier, Amsterdam, 1997).
%
%\bibitem{art:06VB}
%M. Vladkov and J. L. Barrat,
%``Linear and nonlinear viscoelasticity of a model unentangled polymer melt: Molecular dynamics and Rouse mode analysis'',
%Macromol. Theory Simul. {\bf 15}, 252 (2006).
%
%\bibitem{art:86LTI}
%X. L. Luo and R. I. Tanner,
%``A streamline element scheme for solving viscoelastic flow problems. Part I. Differential constitutive equations'',
%J. Non-Newtonian Fluid Mech. {\bf 21}, 179 (1986).
%
%\bibitem{art:86LTII}
%X. L. Luo and R. I. Tanner,
%``A streamline element scheme for solving viscoelastic flow problems. Part II. Integral constitutive models'',
%J. Non-Newtonian Fluid Mech. {\bf 22}, 61 (1986).
%
%\bibitem{art:90RAB}
%D. Rajagopalan, R. A. Brown and R. C. Armstrong,
%``Finite element methods for calculation of steady, viscoelastic flow using constitutive equations with a Newtonian viscosity'',
%J. Non-Newtonian Fluid Mech. {\bf 36}, 159 (1990).
%
%\bibitem{art:95GF}
%R. Gu\'enette and M. Fortin,
%``A new mixed finite element method for computing viscoelastic flows'',
%J. Non-Newtonian Fluid Mech. {\bf 60}, 27 (1995).
%
%\bibitem{art:98B}
%F. P. T. Baaijens,
%``Mixed finite element methods for viscoelastic flow analysis: a review'',
%J. Non-Newtonian Fluid Mech. {\bf 79}, 361 (1998).
%
\bibitem{art:02HKAL}
K. Hyun, S. H. Kim, K. H. Ahn, S. J. Lee,
``Large amplitude oscillatory shear as a way to classify the complex
	fluids'',
J. Non-Newtonian Fluid Mech. {\bf 107}, 51 (2002).
%
\bibitem{art:05CHAL}
K. S. Cho, K. Hyun, K. H. Ahn, and S. J. Lee,
``A geometrical interpretation of large amplitude oscillatory shear
	response'',
J. Rheol. {\bf 49}, 747 (2005).
%
\bibitem{art:08EHM}
R. H. Ewoldt, A. E. Hosoi, and G. H. McKinley,
``New measures for characterizing nonlinear viscoelasticity inlarge
	amplitude oscillatory shear'',
J. Rheol {\bf 52}, 1427 (2008).
%
\bibitem{art:10EM}
R. H. Ewoldt and G. H. McKinley,
``On secondary loops in LAOS via self-intersection of Lissajous-Bowditch
	curves'',
Rheol Acta {\bf 49}, 213 (2010).
\end{thebibliography}
%
%\acknowledgments
%Insert here the text.
%

%%%%%%%%%%%%%%%%%%%%%%%%%%%%%%%%%%%%%
%\begin{comment}
\clearpage
\clearpage
\clearpage
\clearpage
\clearpage
\clearpage
\clearpage
\clearpage
\clearpage
\clearpage
\clearpage
\clearpage
\clearpage
\clearpage
\begin{figure}[htbp]
\includegraphics[scale=1]{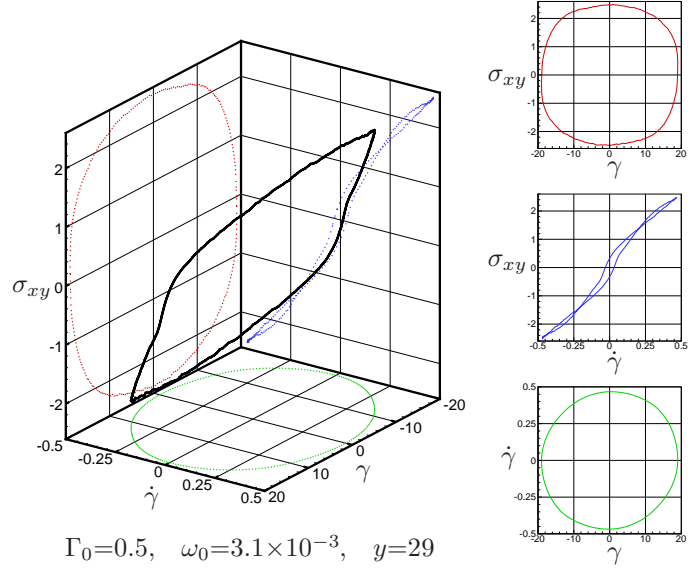}
\caption{(Color)
3D Lissajous-Bowditch curves of the local shear stress $\sigma_{xy}$, local
 strain $\gamma$, and local strain rate $\dot \gamma$ near the
 rapidly oscillating plate. 2D projections onto the planes of the shear
 stress $\sigma_{xy}$ and
 strain $\gamma$, stress $\sigma_{xy}$ and strain rate $\dot\gamma$, and
 strain rate $\dot\gamma $ and strain $\gamma$ are also shown on the right side.
}
\end{figure}
%\end{comment}

\end{document}